\documentclass[pra,twocolumn,showpacs,superscriptaddress,10pt]{revtex4-1}
\usepackage[colorlinks,citecolor=blue,linkcolor=red,dvipdfm]{hyperref}
\usepackage{amsmath,amssymb,graphicx}

\begin{document}

\title{Localized modes and dark solitons sustained by nonlinear defects}

\author{Liangwei Zeng}
\affiliation{College of Physics and Optoelectronic Engineering, Shenzhen University, Shenzhen 518060, China}
\affiliation{Shenzhen Key Laboratory of Micro-Nano Photonic Information Technology, Key Laboratory of Optoelectronic Devices and Systems of Ministry of Education and Guangdong Province, College of Physics and Optoelectronics Engineering, Shenzhen University, Shenzhen 518060, China}

\author{Vladimir V. Konotop}
\affiliation{Departamento de F\'{i}sica and Centro de F\'{i}sica Te\'orica e Computacional, Faculdade de Ci\^encias, Universidade de Lisboa, Campo Grande, Ed. C8, Lisboa 1749-016, Portugal}

\author{Xiaowei Lu}
\affiliation{College of Physics and Optoelectronic Engineering, Shenzhen University, Shenzhen 518060, China}
\affiliation{Shenzhen Key Laboratory of Micro-Nano Photonic Information Technology, Key Laboratory of Optoelectronic Devices and Systems of Ministry of Education and Guangdong Province, College of Physics and Optoelectronics Engineering, Shenzhen University, Shenzhen 518060, China}

\author{Yi Cai}
\affiliation{College of Physics and Optoelectronic Engineering, Shenzhen University, Shenzhen 518060, China}
\affiliation{Shenzhen Key Laboratory of Micro-Nano Photonic Information Technology, Key Laboratory of Optoelectronic Devices and Systems of Ministry of Education and Guangdong Province, College of Physics and Optoelectronics Engineering, Shenzhen University, Shenzhen 518060, China}

\author{Qifan Zhu}
\affiliation{College of Physics and Optoelectronic Engineering, Shenzhen University, Shenzhen 518060, China}
\affiliation{Shenzhen Key Laboratory of Micro-Nano Photonic Information Technology, Key Laboratory of Optoelectronic Devices and Systems of Ministry of Education and Guangdong Province, College of Physics and Optoelectronics Engineering, Shenzhen University, Shenzhen 518060, China}

\author{Jingzhen Li}
\email{\underline{lijz@szu.edu.cn}}
\affiliation{College of Physics and Optoelectronic Engineering, Shenzhen University, Shenzhen 518060, China}
\affiliation{Shenzhen Key Laboratory of Micro-Nano Photonic Information Technology, Key Laboratory of Optoelectronic Devices and Systems of Ministry of Education and Guangdong Province, College of Physics and Optoelectronics Engineering, Shenzhen University, Shenzhen 518060, China}

\begin{abstract}
Dark solitons and localized defect modes against periodic backgrounds are considered in arrays of waveguides with defocusing Kerr nonlinearity constituting a nonlinear lattice. Bright defect modes are supported by local increase of the  nonlinearity, while dark defect modes are supported by a local decrease of the nonlinearity. Dark solitons exist for both types of the defect, although in the case of weak nonlinearity they feature side bright humps making the total energy propagating through the system larger than the energy transferred by the constant background. All considered defect modes are found stable. Dark solitons are characterized by relatively narrow windows of stability. Interactions of unstable dark solitons with bright and dark modes are described.
\end{abstract}

\maketitle

Periodic modulations of parameters and localized defects embedded in a guiding medium are two common tools for manipulating, confining, and directing the light. When guiding media are nonlinear these factors can be introduced as variations of either linear refractive index, or   Kerr coefficients, or both. Nowadays the effect of linear and nonlinear lattices on nonlinear wave propagation is very-well understood and particularly well studied in optical applications~\cite{REVNL} and in the physics of Bose-Einstein condensates~\cite{BECS,BECS1} (see also \cite{Malomed} for specific description of the soliton management). Due to versatility of those tools, the combined effect of linear localized defects and linear lattices has been thoroughly studied too. Focusing on continuous models (for a review on discrete systems see e.g.~\cite{Tsironis,review}),  bright and dark solitons have been described in linear lattices with defects~\cite{Yang2005,BKP1,BKP2,DS2}, in  nonlinear lattices with linear defects \cite{DEF5,DEF6}, in media with periodically varying saturable nonlinearity in the presence of  defects~\cite{Yang2011}, in periodic media with a dissipative~\cite{KKV1} and parity-time symmetric~\cite{PT1,PT2} defects, combined linear and nonlinear lattices~\cite{NZB,NPNL2}, as well as at the interface between dissimilar nonlinear lattices \cite{NPNL1}. Waves in pure nonlinear lattices (without defects) were also studied theoretically~\cite{Berger,Stuart,BECS2,Abdullaev2005,Fibich,Abdullaev2008,NL2,Cuevas,Marek,NL2019} and observed experimentally~\cite{nonlin_latt1}.  In spite of the considerable attention to the effect of the nonlinear lattices, modes in pure nonlinear lattices with a (nonlinear) defect remain unexplored. This determines our main goal, that is description of defect modes sustained by nonlinear defects of the periodically varying defocusing Kerr nonlinearity. The linear properties of the medium are considered homogeneous. We are interested in nonlinear modes sustained by localized defects against nonzero backgrounds. More specifically we describe the shapes and stability of dark solitons, defect modes, as well as coupled states of defects modes and dark solitons propagating in such media.

We consider a paraxial beam polarized along $y$-axis and propagating along $z$ direction. Denoting the dimensionless field amplitude by $E(z,x)$, the model equation can be written as
\begin{equation}
i\frac{\partial E}{\partial z}=-\frac{1}{2}\nabla^2E+\xi(x)\left|E\right|^2E.
\label{NLSE}
\end{equation}
Here $\xi(x)=(1+\Delta(x))\xi_0(x)$
describes periodic modulation of the nonlinearity (a nonlinear lattice). The function $\zeta_0(x)$  models the nonliner lattice, while $\Delta(x)$ describes a localized defect or several defects. It will be assumed that the nonlinearity is either defocusing or eventually can acquire zero values. Respectively, we require $\xi_0(x)\geq 0$, and $\Delta(x)\geq -1$ with $\Delta(x)\to 0$ at $|x|\to\infty$. Thus, $\xi(x)\geq 0$. In the absence of the defect  both the amplitude and the period of the  nonlinear lattice $\xi_0(x)$ can be scaled out. The scaling of the amplitude and of the transverse width of the defect can also be done in the absence of the lattice. When both types of the modulation of the nolinearity are present the spatial scale and the amplitude of one of them is fixed. Below we fix the scales by setting $\xi_0(x)=\cos^2(\pi x)$. For this choice the medium becomes purely linear in the points $x=\pi/2+n$ where $n$ is an integer.

We consider stationary solutions of Eq.~(\ref{NLSE}) $E=\mathcal{E}(x) e^{ibz}$, where $b$ is the propagation constant, satisfying nonzero boundary conditions
\begin{eqnarray}
\label{bound-cond}
 	\lim_{x\to-\infty}\mathcal{E}(x;b)= \sigma \mathcal{E}_0(x;b), \qquad 	\lim_{x\to\infty}\mathcal{E}(x)= \mathcal{E}_0(x;b)
\end{eqnarray}
where $\sigma=\pm 1 $ and $\mathcal{E}_0(x;b)$ is a periodic solution of the equation
 \begin{eqnarray}
 \label{background}
	-\frac{1}{2} \frac{d^2\mathcal{E}_0}{dx^2}+\xi_0(x)\left|\mathcal{E}_0\right|^2\mathcal{E}_0=-b\mathcal{E}_0
\end{eqnarray}
with the period {1: $\mathcal{E}_0(x;b)=\mathcal{E}_0(x+1;b)$.}
One can show that the background solution $\mathcal{E}_0(x;b)$ can be chosen as pure real (or having an unessential constant phase). Since $\xi_0(x)\geq 0$ one readily concludes that solutions $\mathcal{E}_0(x)$ exist only for $b<0$. We also mention that Eq.~(\ref{background}) admits singular $2-$periodic solutions. In particular, for the propagation constant $b=-\pi^2/2$, we obtain $\mathcal{E}_0(x)=\pi/\sin(\pi x)$. The consideration here is restricted only to bounded background solutions with $\mathcal{E}_0(x)>0$. The stability properties of such background solution are established in \cite{Stuart}.

A single defect having a strength $\delta_0$ and width $\ell$, and centered at $x=x_0$, will be modeled by the Heaviside step-function $\theta(x)$: $\Delta(x)=\delta_0\theta\left(\ell/2-|x-x_0|\right)$. Below we address two situations, where $\delta_0>0$ (strong nonlinearity defect) and $\delta_0<0$ (weak nonlinearity defect). In order to  characterize families of nonlinear modes we introduce the quantity
\begin{eqnarray}
	 P=\int (|E|^2-|E_0|^2)dx
\end{eqnarray}	
characterizing "excess" of the $P$ transferred by these modes with respect to the power of the background.

\begin{figure}[tbp]
\begin{center}
\includegraphics[width=1\columnwidth]{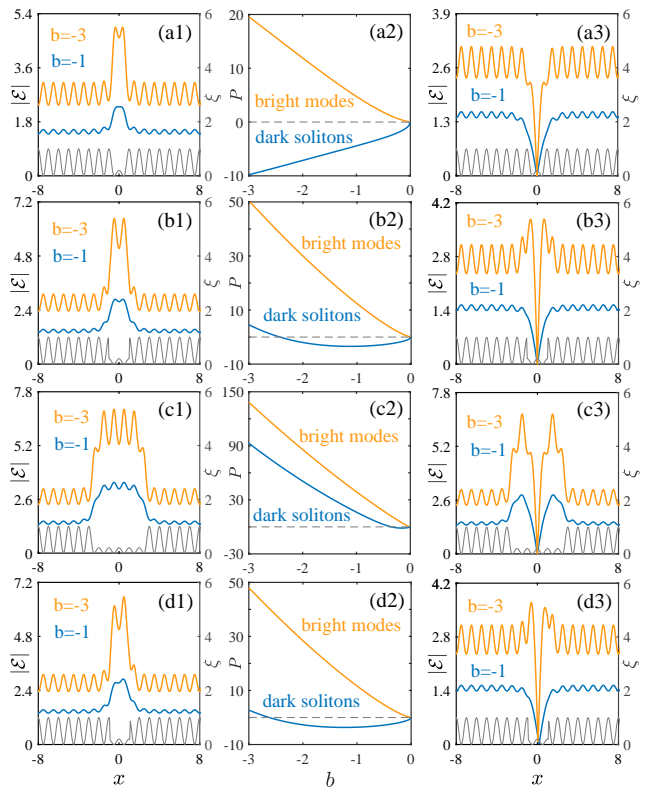}
\end{center}
\caption{(a) Profiles of bright modes (left column) and dark modes (right column) against background with different values of propagation constant $b$ and their power $\Delta P$ {\it versus} $b$ (middle): (a1-a3) with $\ell=1$ at $x_0=0$; (b1-b3) with $\ell=2$ at $x_0=0$; (c1-c3) with $\ell=5$ at $x_0=0$; (d1-d3) with $\ell=1$ at $x_0=0.1$. The gray lines here and below represent the profiles of nonlinearities. $\delta_0=-0.8$ is used for all panels. Dashed lines in the middle column correspond to $P=0$.}
\label{fig1}
\end{figure}

We start by considering local decrease of the defocusig nonlinearity. In Fig.~\ref{fig1} we show results for representative intensity distributions $\mathcal{E}$  of the bright defect modes obtained subject to the boundary condition (\ref{bound-cond}) with $\sigma=1$ (left column) and their families (yellow lines in the middle columns), which are verified to be stable.  The stability of the modes has been verified using both the standard linear stability analysis and the direct propagation with initial perturbation in a form of the random noise (whose amplitude is $1\%$ of the solution amplitude). For linear stability analysis we consider the perturbed field $E=\left[\mathcal{E}(x)+u_+(x) e^{\lambda z}+u_-^*(x)e^{\lambda ^*z}\right]e^{ibz}$, where $u_\pm(x)$ are small perturbations, $\lambda$ is a complex spectral parameter, and asterisk denotes complex conjugation. Respectively we solve the linear eigenvalue problem:
\begin{equation}
i\lambda u_\pm=\mp \frac{1}{2}\nabla^2u_\pm\pm bu\pm\xi(x)\mathcal{E}^2(x)(2u_\pm+u_\mp).
 \label{LAS}
\end{equation}
The solution $\mathcal{E}(x)$ is stable when  $\lambda$ is pure imaginary and unstable otherwise.

We investigated different defect widths $\ell$ and different locations $x_0$ of the defects as illustrated in   Fig.~\ref{fig1}.  Generally speaking, the shape of the defect does not follow exactly the profile of the nonlinearity modulation: in panel (a1) one observes change from the single hump to double hump shapes of   the defect mode when the the absolute value of the propagation constant increases. Thus at sufficiently small $|b|$ the pick intensity is  achieved in the local maximum of the defocusing nonlinearity. In Figs.~\ref{fig1} (b1) and (c1) we show bright defect modes at different widths of the defects. Panel (d1) illustrates a stable asymmetric mode supported by asymmetric defect. Interestingly, at the same value of the propagation constant the asymmetric mode has
appreciably large amplitude and the intensity almost two times bigger than the symmetric mode sustained by the defect of the same width (cf. the modes and intensities on the upper and lower panels of Fig.~\ref{fig1}). All bright modes obtained in the interval $b\in(-10,0)$ were found stable.

In the right column of Fig.~\ref{fig1} we show dark soliton solutions sustained by the weak-nonlinearity defects and obtained subject to boundary conditions (\ref{bound-cond}) with $\sigma=-1$. As one could expect, dark soliton families start with powers $P$  decaying with absolute values of the negative propagation constant (the middle column of Fig.~\ref{fig1}). The slope of $P(b)$ however changes as $|b|$ increases (see in the non-monotonous families of dark solitons in panels (b2), (c2), and (d2) of Fig.~\ref{fig1}) and below some value of $b$ the "excess" power $P$ of a dark mode becomes positive. The increase of $P$ is explained by the change of the soliton shapes with $|b|$: we observe that black solitons feature relatively large energy humps (Figs.~\ref{fig1} (b3) and d(3)). The humps are out of phase, and thus the profile of the dark soliton can be viewed as a dipole soliton against a background. The positive power propagating in the area of the bright humps grows with $|b|$ and at some point it overpasses the negative power excess of the central dark domain. Such solitons are beams with side power picks. When the defect is asymmetric, the asymmetry of the soliton shape is more pronounced at large  $|b|$ [see Fig.~\ref{fig1} (d3)].

\begin{figure}[tbp]
\begin{center}
\includegraphics[width=1\columnwidth]{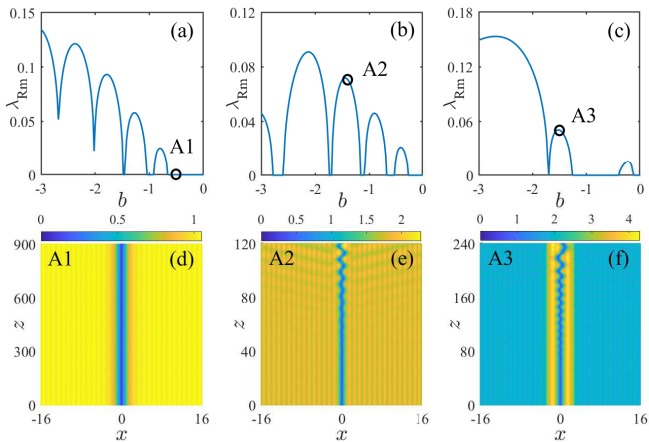}
\end{center}
\caption{The maximal real part of $\lambda$ for dark solitons {\it vs.} $b$ (upper row) and examples of the evolution of the instability of the solutions indicated by points A1 ($b=-0.5$), A2 ($b=-1.4$) and A3 ($b=-1.5$) with different widths of defect at $x=0$: $\ell=1$ (left column), $\ell=2$ (middle column), and $\ell=5$ (right column). In all panels $\delta_0=-0.8$.}
\label{fig2}
\end{figure}

In the upper row of Fig.~\ref{fig2} we show the instability increments of the branches of dark solitons shown in Fig.~\ref{fig1} $\ell=1,2,5$. The distribution of the stability domains depends on the width of the defect. Dark solitons feature some stability domains. The characteristic scenarios of the development of the oscillatory instability of dark solitons are clearly displayed.

\begin{figure}[tbp]
\begin{center}
\includegraphics[width=1\columnwidth]{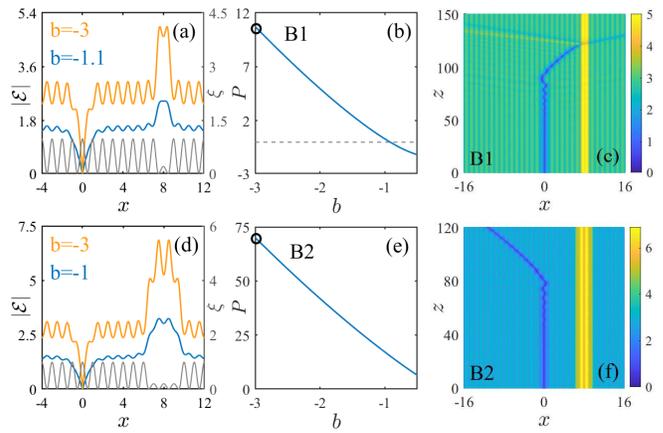}
\end{center}
\caption{Profiles of coupled dark solitons (left column), their power $P$ versus $b$ (middle column) and propagation (right column) with different values of width of defect at $x_0=8$: (a-c) with $\ell=1$; (d-f) with $\ell=3$. The propagation of the solutions marked by B1,B2 in panels (b,e) are presented in panels (e,f) respectively. $\delta_0=-0.8$ is used for all panels. Dashed line in (b) corresponds to $P=0$.}
\label{fig3}
\end{figure}

Now we turn to coupled states of dark solitons and bright defect modes, examples of which are shown in Fig.~\ref{fig3}. One can show that there is a minimal distance between the zero of a dark soliton and the nearest maximum of the defect mode (we call them $x_s$ and $x_m$ respectively). Mathematically, it is obtained from a simple upper bound on the derivative of the field, that follows from (\ref{NLSE}). Recalling  that the field is real, and considering $x_s<x<x_m$ (the  case $x_m<x_s$  is treated similarly) we obtain:
\begin{eqnarray}
\frac{dE}{dx}=2\int_{x_s}^{x}\left[\xi(x) E^3-|b|E\right]dx\leq 2\xi_{\rm max} E^3(x_m)
\end{eqnarray}
where $\xi_{\rm max}=\max_x\xi(x) $, and we used that $E_x(x_s)=0$ as well as the fact that $\max_x E(x)=E(x_m)>0$. Thus
\begin{eqnarray}
x_m-x_s\geq \frac{1}{2\xi_{\rm max}E^2(x_m) }
\end{eqnarray}

\begin{figure}[bp]
\begin{center}
\includegraphics[width=1\columnwidth]{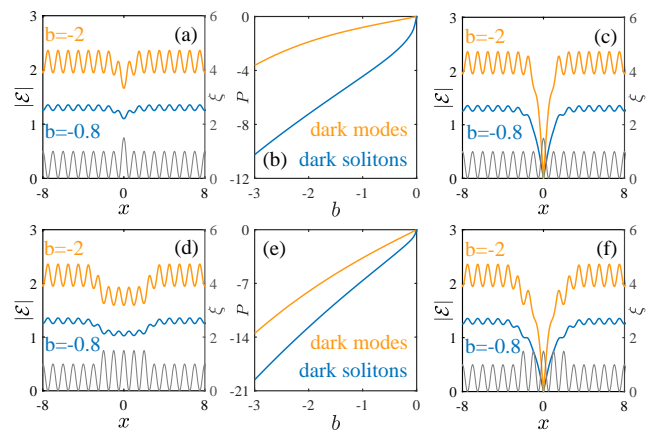}
\end{center}
\caption{Profiles of dark modes (a), (d), of dark solitons (c), (f) and of their families (b), (e) sustained by a strong nonliearity defect cantered as $x_0=0$ of the width $\ell=1$ (upper row) and  $\ell=5$ (bottom row). $\delta_0=0.5$ is used for all panels. }
\label{fig4}
\end{figure}

Due to the presence of a dark soliton, the coupled states bifurcate starting from finite negative value of the propagation constant $b$ with either negative [Fig~\ref{fig3} (b)] or positive [Fig~\ref{fig3} (e)] power depending on the width of the defect (The larger $\ell$, the larger the maximal power of these coupled state). In Figs.~\ref{fig3} (c) and (f) we show unstable dynamics of the coupled dark soliton and bright modes. Being initially perturbed by small random perturbation, a dark soliton after an initial interval of developing instability starts motion either towards the bright mode (the latter being trapped by the local decrease of the nonlinearity) as illustrated in Fig.~\ref{fig3} (c) or outwards the bright modes, as shown in Fig.~\ref{fig3} (f). After interaction with the bright defect mode the dark soliton is split in backward propagating   small-amplitude bright soliton against a background and a small-amplitude dark pulse moving outwards the defect in the positive direction. This type of evolution is illustrated in Fig.~\ref{fig3} (c).

Turning to a strong nonlinearity defect, i.e., to a defect where the nonlinear lattice has local increase of the amplitude, in Fig.~\ref{fig4} we show  representative examples of defect modes (left column)  and dark solitons (right column). As before, the defect modes and solitons are subject to the boundary conditions (\ref{bound-cond}) with $\sigma=+1$ and $\sigma=-1$, respectively.   The respective families of the solutions are shown in the middle column. Now the defect modes represent local decrease of the field amplitude, i.e., they are "dark",  while dark solitons do not feature bright humps at any intensity of the background (cf. the Figs.~\ref{fig4} (c), (f) and Figs.~\ref{fig1} (b3), (c3), and (d3)).

\begin{figure}[tbp]
\begin{center}
\includegraphics[width=1\columnwidth]{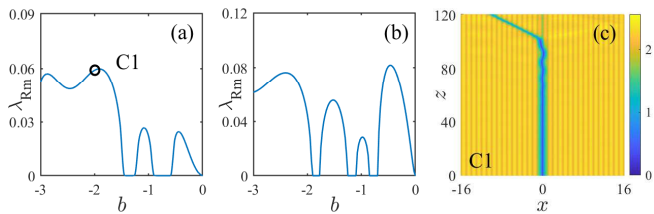}
\end{center}
\caption{The maximal real part of $\lambda$ for dark solitons {\it vs.} $b$ sustained by a strong nonliearity defect cantered as $x_0=0$ of the width $\ell=1$ (a) and  $\ell=5$ (b). (c) Perturbed propagation of the instability of the solutions indicated by points C1 ($b=-2$). The defect strength is $\delta_0=0.5$ is used for all panels.}
\label{fig5}
\end{figure}

Like in the case of weak nonlinearity defect, the dark defect modes are stable for the whole domain shown in Figs.~\ref{fig4} (b) and (e), while the dark soliton is stable only in small intervals of the propagation constant, featuring instability in the small-amplitude (small $b$) limit. Typical example of the evolution of the unstable dark soliton is shown in Fig.~\ref{fig5} (c).

\begin{figure}[tbp]
\begin{center}
\includegraphics[width=1\columnwidth]{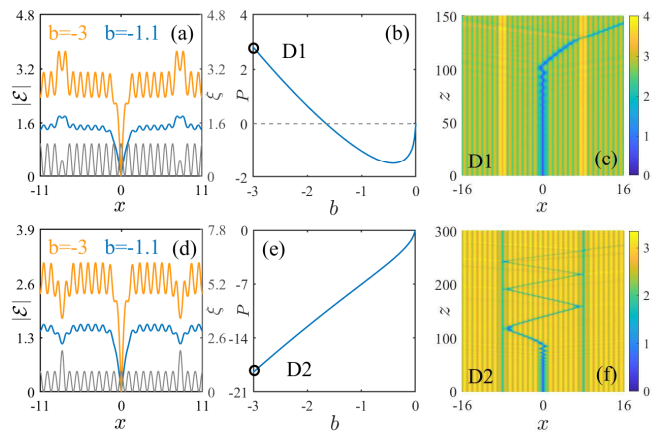}
\end{center}
\caption{Complexes of dark solitons coupled to two defect modes located at $x=\pm8$ in the cases of weak nonlinearity defect at $\delta_0=-0.5$, i.e., bright defect modes (upper row, ) and  of strong nonlinearity defect at $\delta_0=0.5$, i.e., dark defect modes (bottom row).
Characteristic profiles of the complexes (left column), branches (middle column), and  examples of unstable dynamics of a dark soliton (right column) are shown.
 In all panels $\ell=1$ .}
\label{fig6}
\end{figure}

Finally, in Fig.~\ref{fig6} we addressed the cases of nonlinear lattices with two identical defects, of weak nonlinearity (upper row) and strong nonlinearty (bottom row) where a dark soliton is created in the central part having no defect.  Unlike in all above cases, now the dark soliton is centered at a regular lattice site, i.e., at a the local maximum of the nonlinear lattice, exactly in the middle between the defect modes (similar behaviour, not shown here, we also observed for dark solitons centered at the zero at the points where the potential is exactly zero). The upper row illustrates the complex of a dark soliton with two bright defects. The family of such complexes behave similarly to the family of dark solitons on the defect local minimum (cf. Fig.~\ref{fig6} (b) and Figs.~\ref{fig1} (b2), (d2)) although the change of the sign of $P$ now is due to the bright defect modes, while bright side humps observed for a single soliton (see the right column of Fig.~\ref{fig1}) are not observed. Initial perturbation due to oscillatory instability results in motion of a dark soliton to towards one of the defect modes. The dynamics of the interaction of the dark soliton with the mode is the same as observed above (in Fig.~\ref{fig3} (c)).

When an unstable dark soliton is created between two dark defect modes, at sufficiently large nonlinearity it can be reflected by defect modes, thus manifesting trapped oscillatory motion as shown in Fig.~\ref{fig6} (f). After loosing energy  due to radiation, at some instant (approximately at $z=264$ in Fig.~\ref{fig6} (f)), passes through the defect mode.

To conclude we have reported the localized (defect) modes  and dark solitons on the defects of the nonlinear lattices. Such solitons and modes are created against a stable periodic background sustained by the periodically modulated nonlinearity.  While dark solitons and defect modes separately can also be obtained in a homogeneous medium, their coupling, stability and dynamics are governed by the characteristics of the periodical background. The defect modes
are remarkably stable in the whole studied domains, while dark solitons manifest oscillatory instabilities. An unstable dark soliton splits into a grey soliton and a weak bright soliton against a background propagating in opposite directions.

\medskip
\noindent\textbf{Funding.} National Natural Science Foundation of China (61827815, 62075138), Science and Technology Project of Shenzhen (JCYJ20190808121817100, JCYJ20190808164007485, JSGG20191231144201722),  Portuguese Foundation for Science and Technology (FCT) under Contract no. UIDB/00618/2020.




\medskip

\noindent\textbf{Disclosures.} The authors declare no conflicts of interest.

\end{document}